\documentclass[a4paper, fontsize=11pt]{article}

\renewenvironment{abstract}
               {\list{}{\rightmargin\leftmargin}%
                \item[\hspace*{1cm}\small\textbf{Abstract ---}]\relax}
               {\endlist}

\usepackage[top=.8in, bottom=.8in, left=.8in, right=.8in]{geometry}
\usepackage[fleqn]{amsmath}
\usepackage{amssymb}
\usepackage{enumerate}
\usepackage{amsthm}
\usepackage[pdftex]{graphicx}
\usepackage{balance}
\usepackage{fancyhdr}
\usepackage{sidecap}
\usepackage{appendix}
\usepackage{float}
\usepackage{scrextend}
\usepackage{changepage}
\usepackage{subfigure}
\usepackage{endnotes}

\let\footnote=\endnote

\newtheorem*{Postulate*}{}

\begin{document}

\title{\textbf{Addendum to `On the nonreality of the PBR theorem': \\ disproof by generic counterexample}}

\author{Marcoen J.T.F. Cabbolet\\
        \small{\textit{Center for Logic and Philosophy of Science, Vrije Universiteit Brussel\footnote{E-mail: Marcoen.Cabbolet@vub.be}}}}
\date{}

\maketitle
\begin{abstract} \footnotesize The PBR theorem is widely seen as one of the most important no-go theorems in the foundations of quantum mechanics. Recently, in \emph{Found. Phys.} \textbf{53}(3): 64 (2023), it has been argued that there is no reality to the PBR theorem using a pair of bolts as a counterexample. In this addendum we expand on the argument: we disprove the PBR theorem by a generic counterexample, and we put the finger on the precise spot where Pusey, Barrett, and Rudolph have made a tacit assumption that is false.
\end{abstract}

\section{Introduction}\label{sect:intro}
Looking at the view on the quantum state $\psi$, we can distinguish $\psi$-\emph{epistemic} and $\psi$-\emph{ontic} interpretations of quantum mechanics (QM) \cite{Spekkens,Spekkens2,Harrigan}. Paraphrasing the definitions,
\begin{enumerate}[(i)]
  \item an interpretation of QM is $\psi$-\emph{ontic} if there is a one-to-one correspondence between distinct ontic states of a physical system and distinct quantum states of the physical system;
  \item an interpretation of QM is $\psi$-\emph{epistemic} if there is at least one ontic state of a physical system that corresponds to at least two distinct quantum states of the physical system;
\end{enumerate}
The PBR theorem now famously alleges that assuming some preconditions, $\psi$-epistemic interpretations of QM give rise to predictions about the outcome of a measurement that are inconsistent with the predictions of standard ($\psi$-ontic) quantum mechanics \cite{Pusey}. Consequently, the PBR theorem is widely accepted as a \emph{no-go theorem} that rules out $\psi$-epistemic interpretations of QM. Recently, however, it has been argued that there is no reality to the PBR theorem by showing that it fails for an ensemble of bipartite systems made up of two bolts satisfying its preconditions \cite{Cabbolet}.

The purpose of this addendum is to clarify that the PBR theorem is false because it has a generic counterexample. The gist of the argument is as follows. For starters, the PBR theorem is based on the following assumptions:
\begin{itemize}
  \item assumptions $S_1, \ldots, S_n$ for the quantum states of a physical system;
  \item an additional assumptions $O$ and $E$ reflecting the above clauses (i) and (ii), which are respectively to hold in frameworks of a $\psi$-ontic interpretation and a $\psi$-epistemic interpretation of QM;
  \item an assumption guaranteeing that the quantum state $|\ \psi_0\ \rangle$, in which a bipartite system made up of two subsystems is prepared, is one of four quantum states $|\ \psi_1\ \rangle$, $|\ \psi_2\ \rangle$, $|\ \psi_3\ \rangle$, $|\ \psi_4\ \rangle$;
  \item an assumption that for a bipartite system satisfying the previous assumptions a special measurement is possible that projects the quantum state $|\ \psi_0\ \rangle$ on one of four ``PBR states'' $|\ \xi_1\ \rangle$, $|\ \xi_2\ \rangle$, $|\ \xi_3\ \rangle$, $|\ \xi_4\ \rangle$, for which $\langle\ \psi_j\ |\ \xi_j\ \rangle = 0$.
\end{itemize}
The PBR theorem then consists of the inferences
\begin{subequations}
\begin{gather}\label{eq:PBRtheorem}
  S_1, \ldots, S_n, E, |\ \psi_0\ \rangle = |\ \psi_j\ \rangle \vdash P(|\ \psi_j\ \rangle \rightarrow |\ \xi_j\ \rangle) > 0\\
\label{eq:PBRtheorem2}
  S_1, \ldots, S_n, O, |\ \psi_0\ \rangle = |\ \psi_j\ \rangle \vdash P(|\ \psi_j\ \rangle \rightarrow |\ \xi_j\ \rangle) = 0
\end{gather}
\end{subequations}
where $P(|\ \psi_j\ \rangle \rightarrow |\ \xi_j\ \rangle)$ is the probability that the initial state $|\ \psi_j\ \rangle$ upon measurement projects on the PBR state $|\ \xi_j\ \rangle$. So, $\psi$-epistemic QM yields the prediction $P(|\ \psi_j\ \rangle \rightarrow |\ \xi_j\ \rangle) > 0$, while $\psi$-ontic QM yields the prediction $P(|\ \psi_j\ \rangle \rightarrow |\ \xi_j\ \rangle) = 0$: this contrast is the crux of the PBR theorem.

In this paper, however, we will develop a model $M$ of an ensemble of physical systems, such that the aforementioned assumptions $S_1, \ldots, S_n$ and $E$ of the PBR system are valid in the model---this uses the ($\psi$-epistemic) ensemble interpretation of QM. We thus have
\begin{itemize}
  \item $\models_M S_1$
  \item[]\ \ \ $\vdots$
  \item $\models_M S_{n}$
  \item $\models_M E$
  \end{itemize}
We will then show that
\begin{equation}\label{eq:NegationPBRtheorem}
  |\ \psi_0\ \rangle = |\ \psi_j\ \rangle \models_M P(|\ \psi_j\ \rangle \rightarrow |\ \xi_j\ \rangle) = 0
\end{equation}
which means that \emph{even if} the special measurement is possible, there is no way that the bipartite system prepared in the state $|\ \psi_j\ \rangle$ will end up in the PBR state $|\ \xi_j\ \rangle$, cf. Eq. \eqref{eq:PBRtheorem2}. But not only that. We will also show that
\begin{equation}\label{eq:NoCollapse}
\models_M P(|\ \psi_0\ \rangle \rightarrow |\ \xi_1\ \rangle \vee |\ \psi_0\ \rangle \rightarrow |\ \xi_2\ \rangle \vee |\ \psi_0\ \rangle \rightarrow |\ \xi_3\ \rangle \vee |\ \psi_0\ \rangle \rightarrow |\ \xi_4\ \rangle) > 0
\end{equation}
meaning that there are cases in which projection to one of the four PBR states does not occur at all---therefore, there is no such thing as the special measurement assumed to exist in the PBR theorem.

One might object that the present argument is invalid \emph{because} the ensemble interpretation of QM, by which a system ‘has’ observable properties with a definite value in absence of observation, is not widely accepted. That, however, is wrong thinking. The crux is that the PBR theorem alleges that the ensemble interpretation of QM gives rise to predictions inconsistent with standard QM, cf. Eqs. \eqref{eq:PBRtheorem} and \eqref{eq:PBRtheorem2}, while we show that this is not true. This result is \underline{independent} of the number of proponents of the ensemble interpretation of QM, and \underline{independent} of the ratio between the number of proponents of the ensemble interpretation of QM and the number of proponents of standard QM. \\
\ \\
That being said, the remainder of this paper is organised as follows. Sect. \ref{sect:PBR} goes into the details of the PBR theorem. Sect. \ref{sect:counterexample} introduces a generic counterexample to the PBR theorem. Finally, Sect. \ref{sect:flaw} puts the finger on the precise point where Pusey, Barrett, and Rudolph have made a wrong tacit assumption.

\section{Details of the PBR theorem}\label{sect:PBR}

For starters, let's discuss the assumptions for the PBR theorem. First of all, the PBR theorem pertains to a bipartite system, made up of two subsystems with two properties A and B that satisfy the following assumptions:
\begin{enumerate}[(i)]
  \item the property A of a subsystem can have values 0 and 1, with corresponding orthonormal eigenvectors $|\ 0\ \rangle$ and $|\ 1\ \rangle$;
  \item the property B of a subsystem can have values $-$ and +, with corresponding orthonormal eigenvectors $|\ -\ \rangle$ and $|\ +\ \rangle$;
  \item in the basis $\{|\ 0\ \rangle , |\ 1\ \rangle\}$, we have
{\setlength{\mathindent}{0cm}
\begin{subequations}
\begin{gather}\label{eq:+01}
  |\ +\ \rangle = (\ |\ 0\ \rangle + |\ 1\ \rangle)/ \sqrt{2}\\
|\ -\ \rangle = (\ |\ 0\ \rangle - |\ 1\ \rangle)/ \sqrt{2}
\end{gather}
\end{subequations}}
\end{enumerate}
Importantly, on account of \eqref{eq:+01}, for the two possible values of the property A the respective probabilities that a measurement of the value of the property A has the outcome $a$ under the condition that the (sub)system subjected to measurement has the property B with value $b=+$ are
\begin{equation}\label{eq:statistics}
P(a = 0\ |\ b = +) = \langle\ 0\ |\ +\ \rangle^2 = P(a = 1\ |\ b = +) = \langle\ 1\ |\ +\ \rangle^2 = 1/2
\end{equation}
Secondly, the bipartite system is prepared by preparing each of the subsystems \emph{independently} in either the state $|\ 0\ \rangle$ or the state $|\ +\ \rangle$. The four possible initial states $|\ \psi_0\ \rangle$ of the bipartite system are thus
\begin{equation}\label{eq:FourStates}
|\ \psi_1\ \rangle = |\ 0\ \rangle|\ 0\ \rangle, |\ \psi_2\ \rangle = |\ 0\ \rangle|\ +\ \rangle, |\ \psi_3\ \rangle = |\ +\ \rangle|\ 0\ \rangle, |\ \psi_4\ \rangle = |\ +\ \rangle|\ +\ \rangle
\end{equation}
In addition, it is assumed that we can do a special joint measurement on a bipartite system satisfying the preceding assumptions, which projects on one of the following four states, henceforth to be called \emph{the PBR states}, each of which is orthogonal to a possible initial state of the bipartite system:
\begin{equation}\label{eq:PBR states}
\left\{\begin{array}{lll}
  |\ \xi_1\ \rangle = \frac{1}{\surd 2}\left( |\ 0\ \rangle  |\ 1\ \rangle + |\ 1\ \rangle  |\ 0\ \rangle\right) & \bot & |\ 0 \ \rangle|\ 0\ \rangle\\
  |\ \xi_2\ \rangle = \frac{1}{\surd 2}\left(|\ 0\ \rangle  |\ -\ \rangle + |\ 1\ \rangle  |\ +\ \rangle\right) & \bot & |\ 0 \ \rangle|\ +\ \rangle\\
  |\ \xi_3\ \rangle = \frac{1}{\surd 2}\left(|\ +\ \rangle  |\ 1\ \rangle + |\ -\ \rangle  |\ 0\ \rangle\right) & \bot & |\ + \ \rangle|\ 0\ \rangle\\
  |\ \xi_4\ \rangle = \frac{1}{\surd 2}\left(|\ +\ \rangle  |\ -\ \rangle + |\ -\ \rangle  |\ +\ \rangle\right) & \bot & |\ + \ \rangle|\ +\ \rangle
  \end{array}\right.
\end{equation}
In \cite{Pusey}, this is called an \emph{entangled measurement}; its existence is merely assumed.

\begin{SCfigure}[1.5][h!]
\includegraphics[width=0.4\textwidth]{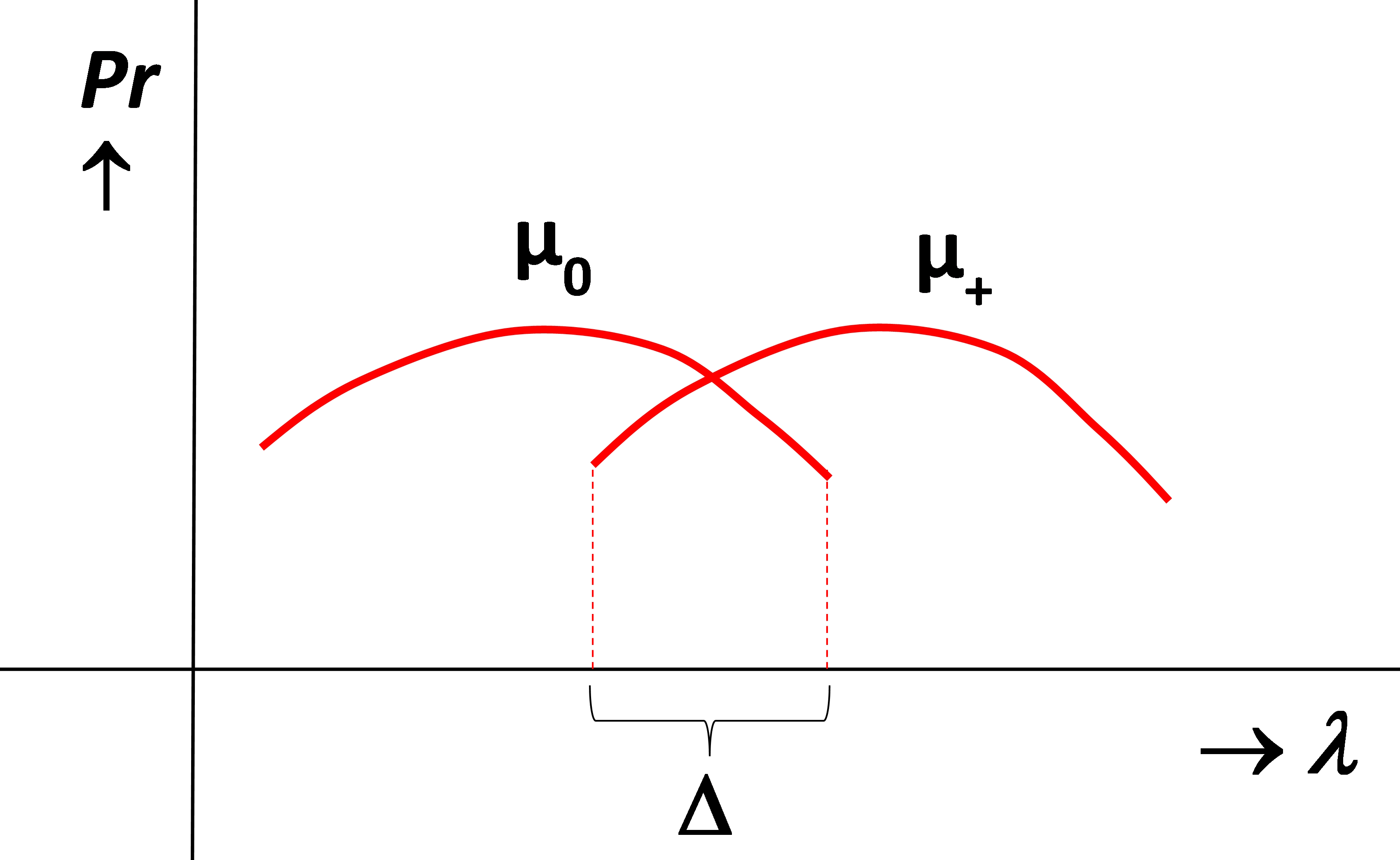}
\caption{Illustration of the overlapping probability distributions $\mu_0(\lambda)$ and $\mu_+(\lambda)$. The overlap region is $\Delta$.}
\label{fig:Probabilities}
\end{SCfigure}

Besides these assumptions, the PBR theorem leans on additional assumptions to get to the inconsistency of the predictions of $\psi$-epistemic QM with the predictions of $\psi$-ontic QM. Namely, \emph{if} the ontic states of a subsystem are labeled by a parameter $\lambda$, \emph{then} for $\psi$-epistemic QM it is assumed that the probability distributions $\mu_0$ and $\mu_+$ for the values 0 and + of the properties A and B of a subsystem as a function of $\lambda$ have an overlap region $\Delta$ as illustrated by Fig. \ref{fig:Probabilities}. For $\psi$-ontic QM, on the other hand, it is assumed that no such overlap region exists.

That being said, the PBR theorem works as follows. On the one hand, \emph{if} we can do the entangled measurement, \emph{then} on account of the orthogonalities in Eq. \eqref{eq:PBR states} standard quantum mechanics (i.e., quantum mechanics with the Copenhagen interpretation) predicts that the measurement will \emph{\textbf{never}} have the outcome $|\ \xi_1\ \rangle$ if the bipartite system has been prepared in the state $|\ 0 \ \rangle|\ 0\ \rangle$, or the outcome $|\ \xi_2\ \rangle$ if the bipartite system has been prepared in the state $|\ 0 \ \rangle|\ +\ \rangle$, or the outcome $|\ \xi_3\ \rangle$ if the bipartite system has been prepared in the state $|\ + \ \rangle|\ 0\ \rangle$, or the outcome $|\ \xi_4\ \rangle$ if the bipartite system has been prepared in the state $|\ + \ \rangle|\ +\ \rangle$.

But on the other hand, we get different predictions in the framework of $\psi$-epistemic QM. Suppose that we repeatedly prepare the bipartite system in any of the four quantum states $|\ \psi_n \ \rangle$ of Eq. \eqref{eq:FourStates}; then there will be a number of cases in which the parameter $\lambda$ for both subsystems is in the overlap region $\Delta$, meaning that both subsystems have property A with value 0 and property B with value +. Then in these cases, the ontic state of the bipartite system corresponds to all four quantum states $|\ \psi_1 \ \rangle, \ldots, |\ \psi_4 \ \rangle$ of Eq. \eqref{eq:FourStates}: while the quantum state $|\ \psi_n \ \rangle$ in which the bipartite system has been prepared is orthogonal to the PBR state $|\ \xi_n \ \rangle$, for the three other quantum states $|\ \psi_k \ \rangle \neq |\ \psi_n \ \rangle$ we have $\langle \psi_k \ |\ \xi_n \ \rangle \neq 0$. Thus speaking, whenever the bipartite system has been prepared in the quantum state $|\ \psi_n \ \rangle$, in these cases the measurement device can ``see'' a quantum state $|\ \psi_k \ \rangle \neq |\ \psi_n \ \rangle$ and project it on the PBR state $|\ \xi_n \ \rangle \bot |\ \psi_n \ \rangle$. That way we arrive at the conclusion that in the framework of $\psi$-epistemic QM a projection $|\ \psi_n \ \rangle \rightarrow |\ \xi_n \ \rangle$ is possible, cf. Eq. \eqref{eq:PBRtheorem}, something that is impossible in the framework of $\psi$-ontic QM, cf. Eq. \eqref{eq:PBRtheorem2}. That's the basic idea behind the PBR theorem.

\section{Generic counterexample to the PBR theorem}\label{sect:counterexample}
Consider an ensemble of systems with properties A and B that can have values $a\in \{0,1\}$ and $b\in \{-,+\}$, respectively, and consider that the ontic state of an individual member of the ensemble is one of four ontic states labeled by $\lambda\in \{1,2,3,4\}$, such that the values $a$ and $b$ of the properties A and B as a function of $\lambda$ are the following:
\begin{subequations}\label{eqs:properties}
\begin{gather}
(\lambda = 1) \Rightarrow (a = 0 \wedge b = -)\\
(\lambda =2) \Rightarrow (a = 0 \wedge b = +)\\
(\lambda =3) \Rightarrow (a = 1 \wedge b = +)\\
(\lambda =4) \Rightarrow (a = 1 \wedge b = -)
\end{gather}
\end{subequations}
So, each of the four ontic states corresponds to two distinct quantum states:
\begin{itemize}
  \item the ontic state with $\lambda =1$ corresponds to $|\ 0\ \rangle$ and $|\ -\ \rangle$
  \item the ontic state with $\lambda =2$ corresponds to $|\ 0\ \rangle$ and $|\ +\ \rangle$
  \item the ontic state with $\lambda =3$ corresponds to $|\ 1\ \rangle$ and $|\ +\ \rangle$
  \item the ontic state with $\lambda =4$ corresponds to $|\ 1\ \rangle$ and $|\ -\ \rangle$
\end{itemize}
By clause (ii) in Sect. \ref{sect:intro}, our model of an ensemble is thus $\psi$-epistemic. The probability distributions $\mu_0$ and $\mu_+$ for the values 0 and + of the properties A and B of a system as a function of $\lambda$ have an overlap region $\Delta = \{2\}$, thereby satisfying the assumption of the PBR theorem.

That being said, if we now prepare a system in the quantum state $|\ 0 \ \rangle$, meaning that we \emph{know} that the property A has value 0, then the parameter $\lambda$ can have any value in the set $\{1,2\}$, being the support of the function $\mu_0$ as dictated by Eqs. \eqref{eqs:properties}. Consequently, if we only know that the property A has value 0, then we do not know the value of the property $B$. That is, the quantum state $|\ 0 \ \rangle$ is \emph{incomplete}. So, if we do $N$ measurements of the value of the property B on a system prepared in the quantum state $|\ 0 \ \rangle$, then the number of times $N(+)$ that the measurement has the outcome $b = +$ is identical to the number of times that the system is in the ontic state with $\lambda = 2$, and the number of times $N(-)$ that the measurement has the outcome $b = -$ is identical to the number of times that the system is in the ontic state with $\lambda=1$. We can now write
\begin{subequations}
\begin{gather}
P(b = +\ |\ a = 0 \wedge \lambda\in\{1,2\}) = \langle\ +\ |\ 0\ \rangle^2 = \lim_{N\rightarrow\infty} \frac{N(+)}{N} = 1/2\\
\label{eq:-|0}
P(b = -\ |\ a = 0\wedge \lambda\in\{1,2\}) = \langle\ -\ |\ 0\ \rangle^2 = \lim_{N\rightarrow\infty} \frac{N(-)}{N} = 1/2
\end{gather}
\end{subequations}
Likewise, for the two possible values $a$ of the property A, the statistical probabilities that a measurement of the value of the property A has the outcome $a$ under the condition that the system has been prepared in the quantum state $|\ + \ \rangle$ satisfy
\begin{subequations}
\begin{gather}
P(a = 0\ |\ b=+\wedge \lambda\in\{2,3\}) = \langle\ 0\ |\ +\ \rangle^2 = \lim_{N\rightarrow\infty} \frac{N(0)}{N} = 1/2\\
\label{eq:1|+}
P(a = 1\ |\ b=+\wedge \lambda\in\{2,3\}) = \langle\ 1\ |\ +\ \rangle^2 = \lim_{N\rightarrow\infty} \frac{N(1)}{N} = 1/2
\end{gather}
\end{subequations}
where $N$ is now the number of measurements of the value of property A on a system prepared in the quantum state $|\ + \ \rangle$. Hence our ensemble of systems satisfies the assumptions of the PBR theorem described in clauses (i)-(iii) in Sect. \ref{sect:PBR}.

What's important is that if we repeatedly, say $N$ times, prepare a system in the quantum state $|\ 0\ \rangle$---that means that we have prepared the system in such a way that we only know that the property A has value 0---then \emph{unbeknownst to us} a distinct number of times $N(+)$ the parameter $\lambda$ will have a value in $\Delta = \{2\}$. But if we then subsequently subject the system to a measurement of the value $a$ of the property A, Eq. \eqref{eq:1|+} does $^*$\textbf{not}$^*$ apply \emph{even though} the system subjected to measurement has property B with value $b=+$. The crux is that the statistical probability in Eq. \eqref{eq:1|+} has been obtained from repeated measurements on an ensemble of systems prepared in the quantum state $|\ +\ \rangle$: for the members of \underline{that} ensemble the parameter $\lambda$ can have any value in the set $\{2,3\}$, and the value $1/2$ of the probability in Eq. \eqref{eq:1|+} is purely due to the value of the number $N(1)$, which is identical to the number of times that the parameter $\lambda$ has the value 3, which is not in $\Delta$. Thus speaking, for the two possible values of property A, the respective probabilities that a measurement of the value of property A has the outcome $a$ under the conditions that the system subjected to measurement has property B with value $b=+$ \emph{and} that the parameter $\lambda$ of the system has a value in $\Delta = \{2\}$ are
\begin{subequations}\label{eq:P}
\begin{gather}\label{eq:P=1}
P(a = 0\ |\ b = + \wedge \lambda \in \Delta) = 1\\
P(a = 1\ |\ b = + \wedge \lambda \in \Delta) = 0 \label{eq:P=0}
\end{gather}
\end{subequations}
Likewise,
\begin{subequations}\label{eq:P2}
\begin{gather}\label{eq:P2=1}
P(b = +\ |\ a = 0 \wedge \lambda \in \Delta) = 1\\
P(b = -\ |\ a = 0 \wedge \lambda \in \Delta) = 0 \label{eq:P2=0}
\end{gather}
\end{subequations}
There is, then, no way that a bipartite system prepared in any of the four quantum states $|\ \psi_n\ \rangle$ of Eq. \eqref{eq:FourStates} can end up in the PBR state $|\ \xi_n\ \rangle \bot |\ \psi_n\ \rangle$ when both subsystems are in the ontic state with $\lambda\in\Delta$.  The nonzero probabilities for a projection $|\ \psi_n\ \rangle \rightarrow |\ \xi_n\ \rangle$ are in each case obtained from inner products $\langle\ 1\ |\ +\ \rangle = 1/\surd2$ or $\langle\ -\ |\ 0\ \rangle = 1/\surd2$, which are factors in an inner product $\langle\ \xi_n\ |\ \psi_k\rangle$ used to calculate the probability of a projection on the PBR state $|\ \xi_n\ \rangle$ when the measurement apparatus ``sees'' the quantum state $|\ \psi_k \ \rangle \neq |\ \psi_n \ \rangle$. Let, for example, the bipartite system be prepared in the quantum state $|\ \psi_1\ \rangle = |\ 0\ \rangle|\ 0\ \rangle$, so that both subsystems have property A with value 0. If in addition $\lambda \in \Delta$ for both subsystems, then both subsystems \emph{also} have property B with value + and the measurement device may then ``see'' the quantum state $|\ \psi_4\ \rangle = |\ +\ \rangle|\ +\ \rangle$. There is then (supposedly) a nonzero possibility $P$ that the bipartite system can be found in a state in which the subsystems have different values of property A, which is the PBR state $|\ \xi_1\ \rangle = \left(|\ 0\ \rangle|\ 1\ \rangle + |\ 1\ \rangle|\ 0\ \rangle\right)/\surd2$:
\begin{equation}
P = \langle\ \xi_1\ |\ \psi_4\rangle^2 = \left(\frac{\langle\ 0\ |\ +\ \rangle\langle\ 1\ |\ +\ \rangle + \langle\ 1\ |\ +\ \rangle\langle\ 0\ |\ +\ \rangle}{\surd2}\right)^2 > 0
\end{equation}
However, as shown by Eqs. \eqref{eq:-|0} and \eqref{eq:1|+}, these inner products $\langle\ -\ |\ 0\ \rangle$ or $\langle\ 1\ |\ +\ \rangle$ reflect statistical probabilities obtained from measurements on an ensemble in which the parameter $\lambda$ varies over a range. If, on the other hand, $\lambda$ is confined to $\Delta$, then we obtain completely different statistical probabilities---namely those given by Eqs. \eqref{eq:P=0} and \eqref{eq:P2=0}. So, in our model $M$, if a bipartite system is initially prepared in a quantum state $|\ \psi_n\ \rangle$, then the probability $P(|\ \psi_n\ \rangle\rightarrow |\ \xi_n\ \rangle)$ that this state upon a measurement projects onto the PBR state $|\ \xi_n\ \rangle$ is given by
\begin{equation}
|\ \psi_0\ \rangle = |\ \psi_n\ \rangle \ \models_M\  P(|\ \psi_n\ \rangle\rightarrow |\ \xi_n\ \rangle) = 0
\end{equation}
\emph{regardless} of how we design the measurement, cf. Eq. \eqref{eq:NegationPBRtheorem}. This invalidates the PBR theorem, since it shows that the prediction $P(|\ \psi_n\ \rangle\rightarrow |\ \xi_n\ \rangle) > 0$ of the PBR theorem for $\psi$-epistemic models of QM, cf. Eq. \eqref{eq:PBRtheorem}, is not valid for our model $M$.

But that's not all. It is true, as mentioned in \cite{Pusey}, that if we consider an infinite ensemble of bipartite systems made up of two subsystems prepared in either the state $|\ 0 \ \rangle$ or in the state $|\ + \ \rangle$, then all four quantum states $|\ \psi_1\ \rangle = |\ 0\ \rangle|\ 0\ \rangle$, $|\ \psi_2\ \rangle = |\ 0\ \rangle|\ +\ \rangle$, $|\ \psi_3\ \rangle = |\ +\ \rangle|\ 0\ \rangle$, $|\ \psi_4\ \rangle = |\ +\ \rangle|\ +\ \rangle$ can be associated to the bipartite system whenever both subsystems will be in an ontic state with $\lambda = 2$, meaning that both subsystems have property A with value 0 and property B with value +. But it is not true that we \emph{therefore} can get a projection $|\ \psi_k\ \rangle\rightarrow |\ \xi_k\ \rangle$, with $|\ \psi_k\ \rangle\bot |\ \xi_k\ \rangle$, when the bipartite system has been prepared in the quantum state $|\ \psi_k\ \rangle$, which is what the PBR theorem alleges. In our model, the opposite is the case: if both subsystems have property A with value 0 and property B with value +, then \emph{none} of the combination of properties of the subsystems entailed by any of the PBR states $|\ \xi_k\ \rangle$ occurs. In other words: if both subsystems have property A with value 0 and property B with value +, a projection on one of the four PBR states does not take place at all, cf. Eq. \eqref{eq:NoCollapse}. Ergo, the assumption that an entangled measurement exists that (always) projects on the PBR states is not true.

\section{Fatal flaw underlying the PBR theorem}\label{sect:flaw}

The point where it goes wrong for Pusey, Barrett, and Rudolph is their tacit assumption that the probabilities of Eq. \eqref{eq:statistics} apply in the context of the ensemble interpretation of QM to any individual member of an ensemble of systems in the quantum state $|\ +\ \rangle$ \emph{regardless of} its ontic state, that is, \emph{regardless of} the value of the parameter $\lambda$. To put that differently: the fatal flaw is the tacit assumption that for every member of an ensemble of systems having property B with value + there is an \emph{intrinsic probability} of 50\% that a measurement of the value of property A will have the outcome $a=1$. It is, namely, \emph{by that tacit assumption} that for the bipartite system prepared in the quantum state $|\ 0\ \rangle|\ 0\ \rangle$, meaning that we \emph{know} that both subsystems have property A with value 0, the measurement can ``flip'' the value of property A of one of the subsystems to 1 when both subsystems also have property B with value +, so that the bipartite system upon measurement ends up in the PBR state $|\ \xi_1\ \rangle = \left(|\ 0\ \rangle|\ 1\ \rangle + |\ 1\ \rangle|\ 0\ \rangle\right)/\surd2$, meaning that we \emph{know} that the two subsystems have different values of property A. By the same mechanism, the bipartite system prepared in any of the other three quantum states $|\ \psi_k\ \rangle$ of Eq. \eqref{eq:FourStates} can end up in a PBR state $|\ \xi_k\ \rangle$ orthogonal to $|\ \psi_k\ \rangle$.

However, this is wrong thinking: while the notion of intrinsic probability is applicable to the framework of \emph{fundamentally probabilistic} Copenhagen quantum mechanics, it is not applicable to a \emph{deterministic} theoretical framework. In the framework of the ensemble interpretation of QM, the probabilities in Eq. \eqref{eq:statistics} are \emph{statistical}, and are obtained from repeated measurements on an \emph{infinite ensemble} of systems in the quantum state $|\ +\ \rangle$: in this ensemble, the parameter $\lambda$ takes all values in its range (\emph{in casu} the set $\{2,3\}$). But suppose we have \emph{an individual system} prepared in the quantum state $|\ 0\ \rangle$, and suppose the parameter $\lambda$ has a value in the overlap region $\Delta$; then it is true that the system to be measured (also) has the property B with value +, but the probability that a measurement of the property A yields the value 1 is 0\% as in Eq. \eqref{eq:P=1}---that is to say, the probabilities in Eq. \eqref{eq:statistics} are not \emph{intrinsic} to having the property B with value +, so for a bipartite system prepared in the quantum state $|\ 0\ \rangle|\ 0\ \rangle$ there is no mechanism by which the value of the property A of one of the subsystems can ``flip'' so that the system ends up in the quantum state $|\ \xi_1\ \rangle$.

Suppose the values of the properties A and B depend on $\lambda$ as in Eqs. \eqref{eqs:properties}; if we then could prepare an ensemble of subsystems of the bipartite systems with $\lambda = 2$, meaning that for each element of the ensemble we \emph{know} that it has the property A with value 0 and the property B with value +, and we would repeatedly measure the value of the property A, we would statistically obtain the probabilities of Eqs. \eqref{eq:P=0} and \eqref{eq:P=1} \emph{instead of} the probabilities of \eqref{eq:statistics}. The PBR theorem, however, leans on the condition that we prepare an ensemble of bipartite systems of which the subsystems are in a quantum state $|\ 0\ \rangle$ or $|\ +\ \rangle$: these quantum states are then \emph{incomplete descriptions} of the state of the system. The probabilities in Eq. \eqref{eq:statistics} are then a reflection of our ignorance: God does not play dice!

\end{document}